\PassOptionsToPackage{bookmarks=false}{hyperref}

\documentclass[
    sigconf,
]{acmart}

\RequirePackage[utf8]{inputenc}     
\RequirePackage[english]{babel}     
\RequirePackage{microtype}          
\RequirePackage{graphicx}           
\RequirePackage[caption=false,font=footnotesize]{subfig}
\RequirePackage{listings}
\RequirePackage{url}
\RequirePackage{fancyref}

\usepackage{url}
\usepackage{xcolor}
\colorlet{punct}{red!60!black}
\colorlet{quote}{orange!60!black}
\definecolor{background}{HTML}{EEEEEE}
\definecolor{delim}{RGB}{20,105,176}
\colorlet{numb}{magenta!60!black}
\lstdefinelanguage{json}{
    basicstyle=\linespread{0.9}\ttfamily\footnotesize,
    numbers=left,
    numberstyle=\scriptsize,
    stepnumber=1,
    numbersep=8pt,
    showstringspaces=false,
    breaklines=true,
    backgroundcolor=\color{background},
    literate=
      {"}{{{\color{quote}{"}}}}{1}
      {:}{{{\color{punct}{:}}}}{1}
      {,}{{{\color{punct}{,}}}}{1}
      {\{}{{{\color{delim}{\{}}}}{1}
      {\}}{{{\color{delim}{\}}}}}{1}
      {[}{{{\color{delim}{[}}}}{1}
      {]}{{{\color{delim}{]}}}}{1},
}
\lstdefinestyle{json}{
  float,
  floatplacement=t,
  belowcaptionskip=-12pt,
  language=json, 
  float=h, 
  numbers=left, 
  numberstyle=\tiny, 
  stepnumber=1, 
  numbersep=1pt, 
  captionpos=b, 
  xleftmargin=1em,
  framexleftmargin=1em
}

\def\sectionname{Section}
\providecommand\figureref[1]{\hyperref[#1]{\figurename~\ref*{#1}}}

\providecommand\tableref[1]{\hyperref[#1]{\tablename~\ref*{#1}}}

\providecommand\listingref[1]{\hyperref[#1]{\lstlistingname~\ref*{#1}}}

\providecommand\sectionref[1]{\hyperref[#1]{\sectionname~\ref*{#1}}}
\providecommand\sectionpageref[1]{\hyperref[#1]{\sectionname~\ref*{#1}~(p.~\pageref{#1})}}

\makeatletter

\@ACM@printacmreffalse
\@ACM@affiliation@obeypunctuationfalse

\@printpermissionfalse
\@printcopyrightfalse

\makeatother

\RequirePackage{caption}   
\RequirePackage{enumitem}  
\graphicspath{ {images/} }

\renewcommand{\baselinestretch}{0.897}  

\AtBeginDocument{

\captionsetup{font={sf,small}}
\captionsetup[figure]{belowskip=-12pt, aboveskip=6pt}

}
\begin{document}%
\hyphenation{WoTify}
\sloppy

\title{WoTify: A platform to bring Web of Things to your devices}


\author{Ege Korkan \vspace{-0.9em}} 
\orcidi{0000-0003-4910-4962}    
\email{ege.korkan@tum.de}%
\affiliation{
  \institution{Technical University of Munich}%
  \country{Germany}
}

\author{Hassib Belhaj Hassine}%
\orcidii{0000-0003-3195-3739}   
\email{hassib.belhaj@tum.de}
\affiliation{
  \institution{Technical University of Munich}%
  \country{Germany}
}

\author{Verena Eileen Schlott}%
\orcidiii{0000-0002-7777-6291}   
\email{verena.schlott@campus.lmu.de}
\affiliation{
  \institution{Ludwig Maximilian University of Munich}%
  \country{Germany}
}

\author{Sebastian Käbisch}%
\orcidiv{0000-0002-0544-4204}   
\email{sebastian.kaebisch@siemens.com}
\affiliation{
  \institution{Siemens AG}%
  \country{Germany}
}

\author{Sebastian Steinhorst}%
\orcidv{0000-0002-4096-2584}   
\email{sebastian.steinhorst@tum.de}
\affiliation{
  \institution{Technical University of Munich}
  \country{Germany}
}

\renewcommand{\shortauthors}{\small E. Korkan, H. Belhaj, V. Schlott, S. Käbisch and S. Steinhorst}

\begin{abstract}%

The Internet of Things (IoT) has already taken off, together with many Web of Things (WoT) off-the-shelf devices, such as Philips Hue lights and platforms such as Azure IoT. 
These devices and platforms define their own way of describing the interactions with the devices and do not support the recently published WoT standards by World Wide Web Consortium (W3C). 
On the other hand, many hardware components that are popular in developer and maker communities lack a programming language independent platform to integrate these components into the WoT, similar to \emph{npm} and \emph{pip} for software packages.
To solve these problems and nurture the adoption of the W3C WoT, in this paper, we propose a platform to WoTify either existing hardware by downloading new software in them or already existing IoT and WoT devices by describing them with a Thing Description.

\end{abstract}%

%
%




\firstpageacks{}


\copyrightyear{2019}
\acmYear{2019}
\setcopyright{none}
\acmConference{\small Second W3C Workshop on the Web of Things}{3-5 June 2019}{Munich, Germany}
\acmDOI{none}
\acmISBN{none}

\maketitle%

%
\section{Introduction}\label{sec:intro}

In the recent years, the Internet of Things\ (IoT) was the focus of many analyses with billions of devices expected to be connected to the Internet.
Some of these billion devices are already out there, connected to the Internet, providing sensing and actuation capabilities. 
More are waiting to be connected to the Internet and thus be part of the IoT.

When we look at the Internet, most of the applications run on the Web layer.
This has resulted on the Web having one of the biggest developer communities behind \cite{stackoverflowreport} and we can foresee the same happening for the Web layer of IoT that was named the Web of Things (WoT).
The WoT in itself is a concept of having Web technologies for IoT devices and offer the same ease of application development composed of IoT devices.
WoT differentiates itself from the Web by being related to underlying hardware such as sensors, controllers, robots, actuators and more, something that was foreign to the Web community.

As with all the Web standards, the Web of Things is also being standardized by the World Wide Web Consortium (W3C) by means of different building blocks such as Thing Description (TD)\cite{wotTD}, Scripting API\cite{scripting_api} and more\footnote{The entire work of the working group can be found at \url{https://www.w3.org/WoT/WG/}}.
This standardization effort has started on 2016 \cite{RaggettDave2016} and the Thing Description standard will be released by the end of 2019.
Being this recent, there has not been enough time to build a community around the W3C WoT, which is important for the acceptance and dissemination of the standard.

When we look at the wider scope of the IoT, there are multiple proposals of IoT marketplaces \cite{Broring2017}\cite{krishnamachari2018i3}. 
These marketplaces offer a way to obtain applications for a wide variety of IoT devices and contribute to the IoT landscape with a convenient and standardized way to obtain services and applications from the IoT devices.
However, these marketplace concepts are not specifically aimed at the W3C WoT applications and devices.

\begin{figure}[t]
  \includegraphics[width=0.45\textwidth]{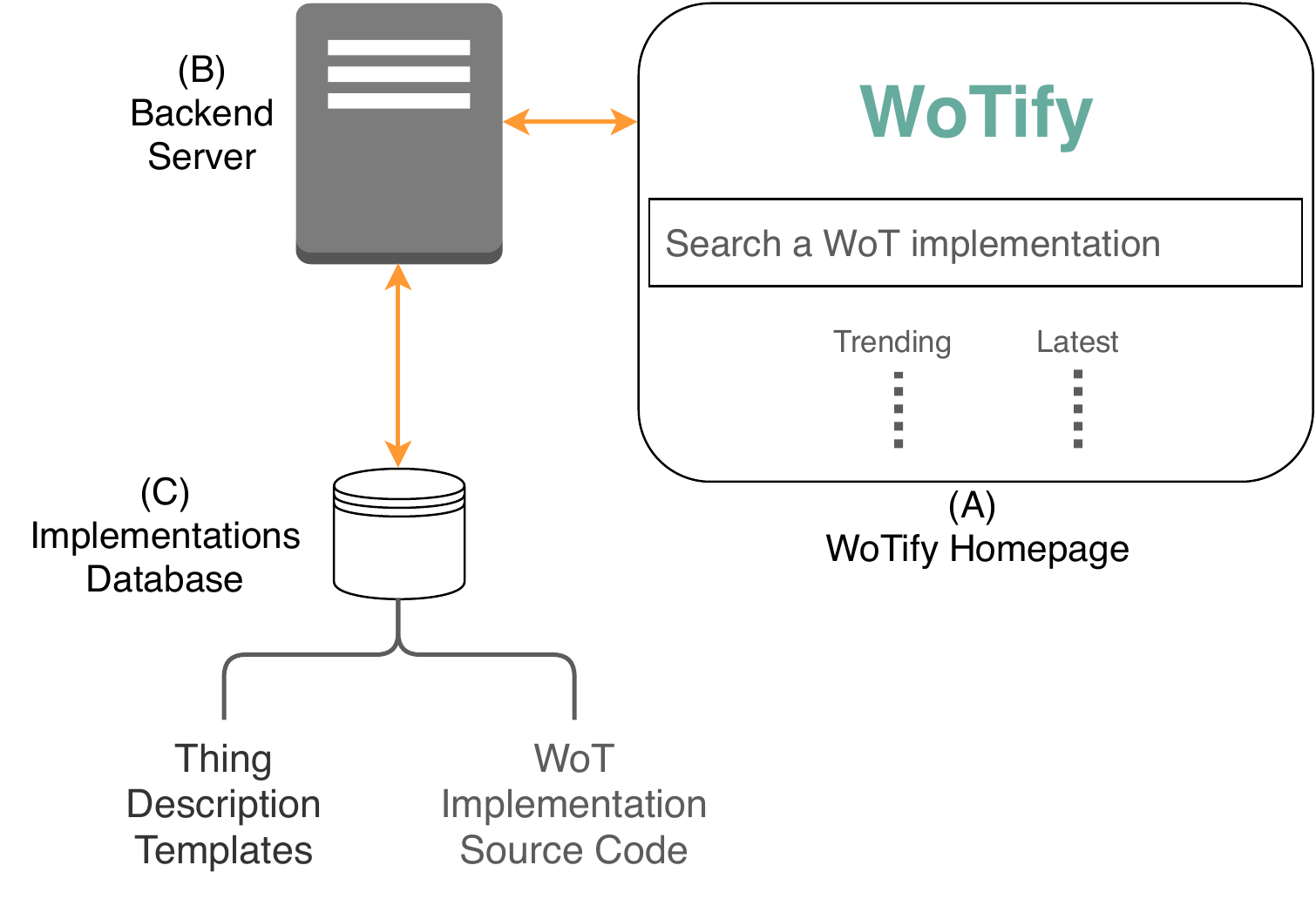}
  \centering
  \caption{WoTify concept illustrating different components of the WoTify system}
  \label{fig:wot-concept}
\end{figure}

  \subsection{Problem Statement}

  For many communities, a dedicated platform is where new members of the community would come to. 
  Even though code repositories like \emph{GitHub}\footnote{\url{https://github.com/}} can host projects, tutorials and more, they remain generic.
  Thus, specific platforms gain importance within communities, with package managers playing a similar role for software specific ones.
  It is possible to see platforms such as \emph{Thingiverse}\footnote{\url{http://thingiverse.com}} for 3D Printer community, \emph{Make:}\footnote{\url{https://makezine.com/projects}} and \emph{Instructables}\footnote{\url{https://www.instructables.com}} for the maker community, \emph{hackster.io}\footnote{\url{https://www.hackster.io}} for hardware design community.
  Another example is the \emph{Node-RED Library}\cite{noderedlibrary} which supports over 3000 projects that are built specifically for the \emph{Node-RED} programming tool.
  Moreover, package repositories and their associated command line interfaces as package managers became the de facto standard for installing, publishing and maintaining software projects, such as \emph{npm}\footnote{\url{http://npmjs.com}} for JavaScript, \emph{pip}\footnote{\url{http://pypi.org}} for Python or \emph{Maven}\footnote{\url{https://maven.apache.org/}} for Java.

  In the case of the W3C WoT, unfortunately there is no such platform that would enable people to find projects or anyone to publish projects in a W3C WoT compatible way, making it tedious to integrate various IoT devices into the WoT.
  After the publication of the Thing Description standard, there will be a need to welcome new members with a platform that consists of projects and the community interaction that is found in other platforms.

  The Eclipse Thingweb project~\cite{eclipsethingweb} aims to address some of these issues but there is currently no repository to group all the existing projects of the WoT community or to allow new projects to be published.
  There are numerous platforms and repositories but they are mostly package managers for a specific programming language and they are not based on WoT technologies like the TD.\@
  This makes it more difficult to convince how diverse the applications of the WoT are, hinders the further acceptance of the WoT and can result in existing IoT device owners to reinvent the wheel, which is the main reason why we often see siloed IoT solutions.

  \subsection{Contributions}

  We propose a platform called WoTify that is able to host W3C WoT projects provided by the community which can be used by anyone to turn their Internet connected devices into W3C WoT compatible devices, i.e.\ to WoTify them\footnote{By WoTify, we mean two things: the process of integrating a device into WoT and the name of the platform that will help with that.}.
  WoTify hosts project pages that can be composed of source code of any programming language or TD templates that can be used to describe already existing closed source and brownfield IoT devices with a TD.\@
  In the end, our platform allows anyone to search for W3C WoT projects or to contribute to the W3C WoT by sharing new implementations.

  In particular, we propose:
  
  \begin{enumerate}

    \item a currently running platform with already existing implementations, ready for the WoT community,
    \item a method to bring WoT functionality to any Internet connected device,
    \item a command line interface idea for the W3C WoT that is independent from a programming language.

  \end{enumerate}

  We imagine WoTify to be the major WoT platform where even manufacturers can find a TD for their device that is provided by the community. 

\section{W3C Web of Things}\label{sec:wot}

The Web of Things is a set of design and programming norms that enable real world internet connected devices to be part of the World Wide Web. 
It started as an academic initiative to with the main goal of enabling device interoperability.
It is based on the use of existing and widespread Web concepts, standards and protocols, such as REST, HTTP, CoAP and JSON, to enable inter device communication and device access.

For example, some Web of Things devices are the Philips Hue lights, even though they might not be marketed as such. 
They can be accessed and controlled using a RESTful API and standard communication protocols such as HTTPS.\@
The problem with such devices is that discovering and understanding how to use their API is not a straightforward task.
The Hue API is described in proprietary websites and datasheets. 
This requires developers to do more work to understand the proprietary API descriptions of each device they are trying to work with.

\begin{figure}[t]
  \includegraphics[width=0.45\textwidth]{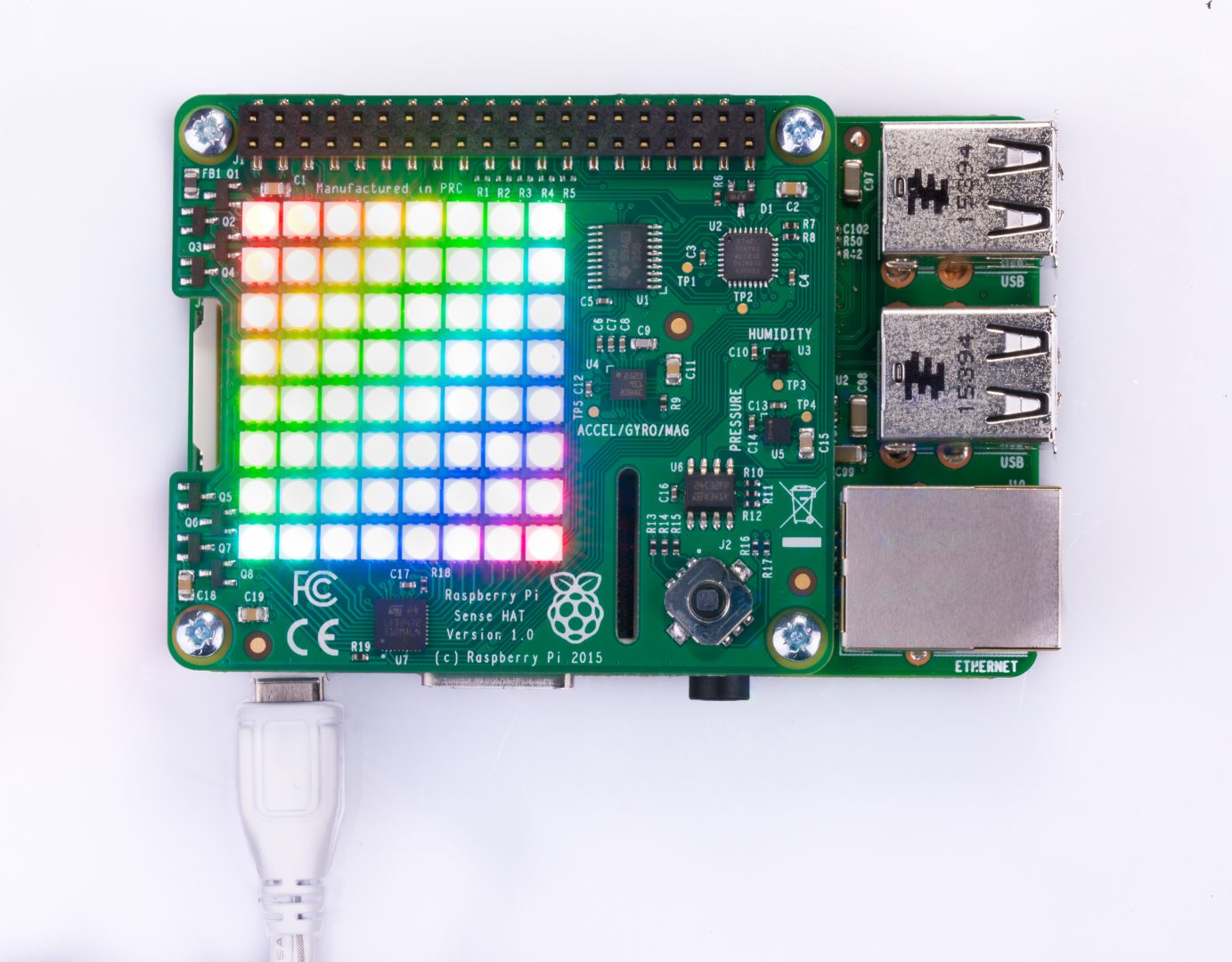}
  \centering
  \caption{Photo of a Raspberry Pi with a Sense HAT attached on it\cite{Raspberry:sensehat}.}
  \label{fig:sensehat}
\end{figure}

To solve this problem, the W3C started the WoT working group in 2016 to standardize a set of mechanisms for describing and working with Web of Things devices with the goal of enabling interoperability between devices, independent from the underlying framework and implementation.
The main building blocks of the W3C WoT standardization effort are now being finished and include the WoT Thing Description\cite{wotTD} and WoT Architecture\cite{wotArch}. 
Of these two, the WoT Thing Description is the primary and most important building block, describing the public interface of a Thing.
A Thing's TD provides a formal description of the functionalities of the Thing and how to use them. 
This removes the need to understand a multitude of non-standard, manufacturer specific API descriptions when integrating devices from different manufacturers.
Another advantage provided by TDs is that they can be written for existing Web connected devices. 
An existing Thing can be made W3C WoT compatible simply by having a TD added to it.

Coming back to the Philips Hue light example discussed above, we can make it W3C WoT compatible simply be creating a TD to describe its API.\@ 
This can be a simple translation of the API description provided on the Philips website to the standard TD format, making the Hue's API understandable to any developer familiar with the WoT TD standard. 
This can enable much quicker development times when integrating it with other WoT enabled devices from other manufacturers, as developers no longer need to navigate multiple manufacturer websites to fetch the API information for each device independently.


Similarly, any Internet connected device can be WoTified.
WoTifying a device can be as trivial as simply writing a TD to describe an existing device's API or as complicated as extending a non-Internet-connected device with network interface to be remotely accessed (using an ESP8266 or a Raspberry Pi for example) and describing the resulting interface in a TD.\@
Any device that can have a Web server with any Web protocol can thus be turned into a W3C WoT enabled device.

An example of a device that may be WoTified this way is the Raspberry Pi Sense HAT.\@
It is a suite of sensors and an LED display that can be attached to a Raspberry Pi like in Figure~\ref{fig:sensehat}. 
Out of the box, the Sense HAT is not an IoT device. 
However, it is possible to write a Web server that runs on the underlying Raspberry Pi and enable the functionality of Sense HAT as a Web service.
Once this work has been done, and a TD describing the resulting API is created, anyone can install this Web server on their Sense HAT attached Raspberry Pi and have it automatically turned into a W3C WoT device.

\section{Concept}\label{sec:concept}


The idea behind WoTify is to have a platform available for the entire WoT community, composed of expert and new members alike.
The community would contribute to WoTify by sharing their WoT implementations or by downloading what the other members of the community have already shared.
Since the word \textit{platform} is used in many different contexts, we want to explain what exactly WoTify can do in this section.


\begin{figure}[t]
  \vspace{-2mm}
  \subfloat[The homepage of Node-RED Library, showing different contributions by the Node-RED community\cite{noderedlibrary}.]{
    \includegraphics[width=0.45\textwidth]{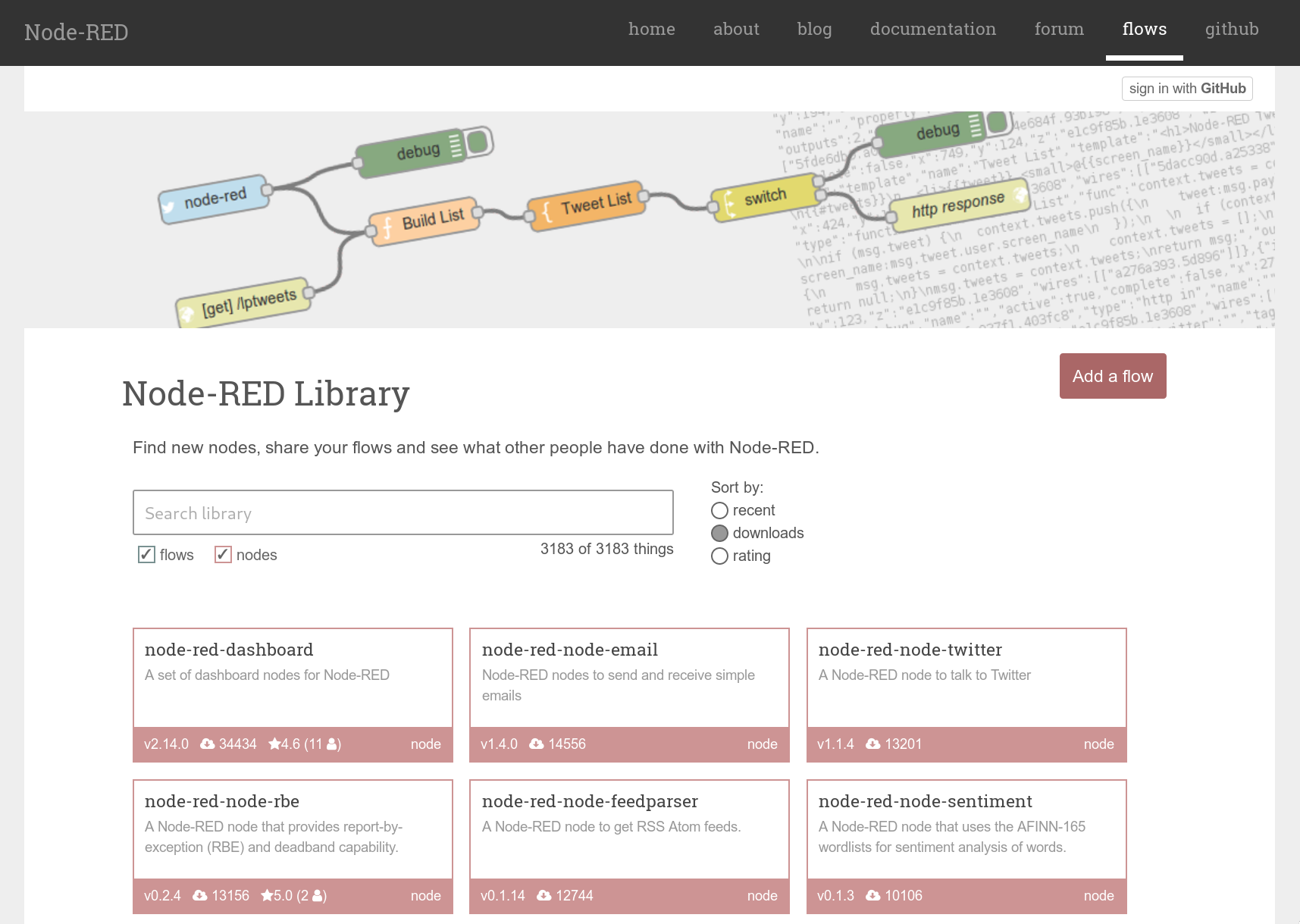}
    \label{fig:node_red}
  }

  \subfloat[A project page of Node-RED Library, containing information to integrate the project into the Node-RED programming tool\cite{nodereddashboard}.]{
    \includegraphics[width=0.45\textwidth]{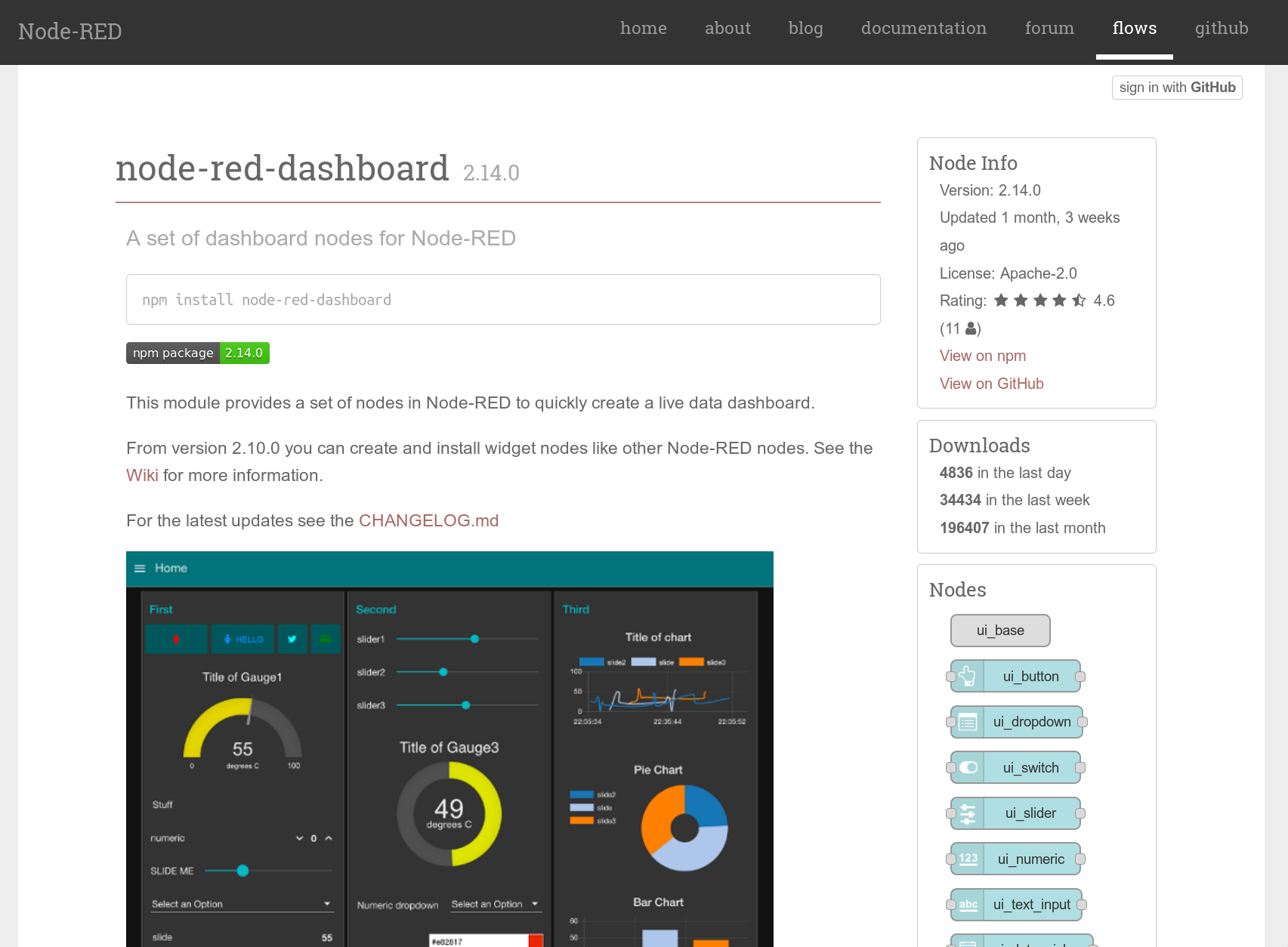}
    \label{fig:node-red-dashboard}
  }

  \caption{Two screenshots of the Node-RED Library platform that can host different projects from contributors}
\end{figure}

We think that Node-RED library is a very good example of a similar platform that has also influenced design decisions of WoTify. 
As seen in Figure~\ref{fig:node_red}, Node-RED library allows anyone to contribute and to search for contributions.
After choosing a contribution, i.e.\ a node, one would be met with a project page like in Figure~\ref{fig:node-red-dashboard}, that contains all the relevant information and statistics about the project.

Our aim with WoTify is to provide the same benefits but in the context of W3C WoT.
This means, being able to search for WoT projects, download them and add new projects.
However, the WoT is not a programming tool like Node-RED.\@
The WoT is a programming language, tool or environment independent way to connect devices on the application layer and relies on standards from the W3C.

Since it relies on a set of standards, in order for a device to count as a W3C WoT device, it should have a TD representation.
A W3C WoT device could be programmed in any programming language but its capabilities and services must be described with a TD.\@
This means that WoTify must support TDs and be able to host projects programmed in any programming language.
Furthermore, one can retrofit already existing IoT devices with a TD and turn them into W3C WoT devices as explained in Section~\ref{sec:wot}.
This means that {WoTify} should also support distribution of TDs of devices without the software that would run on the devices since already existing IoT devices may not allow a brand new software downloaded through WoTify to run on themselves.

\begin{figure}[t]
  \includegraphics[width=0.45\textwidth]{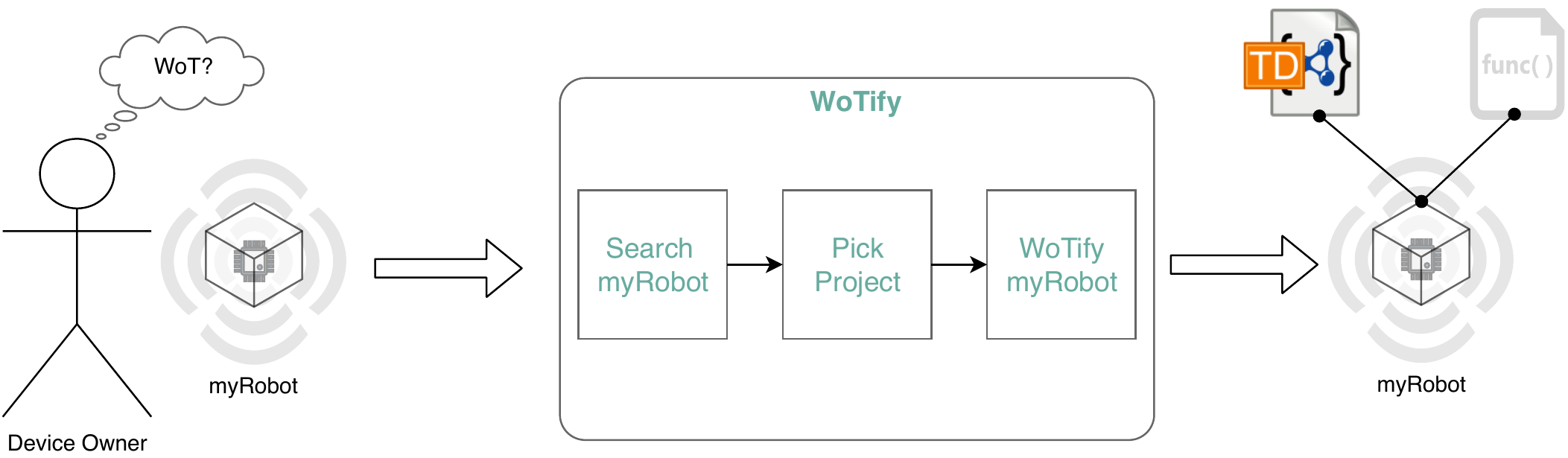}

 \centering
   \caption{How to search for a project to turn an Internet connected device into a Web of Things device}
   \label{fig:searchFlow}
  \end{figure}

In the end, WoTify aims to facilitate the two following aspects:

\textbf{Using a WoT Project:} Once one has a device to integrate into the WoT, may it be a barebone computer like a Raspberry Pi or an ESP8266 with some sensors or an off-the-shelf Philips Hue device, WoTify is the next step on integrating this device into the WoT. As shown in Figure~\ref{fig:searchFlow}, the device owner:

\begin{enumerate}
  \item searches for a WoT device project by providing the device name,
  \item chooses one of the results presented by WoTify based on programming language preference, complexity etc.
  \item if the project is a TD: Downloads the TD and start interacting with the device according to the WoT interaction patterns
  \item if the project is a software implementation: Reads the project page for installation instructions, install the project on the device and start interacting with the device according to the WoT interaction patterns
  \item gives feedback on the project.
\end{enumerate}

\begin{figure}[t]
  \includegraphics[width=0.45\textwidth]{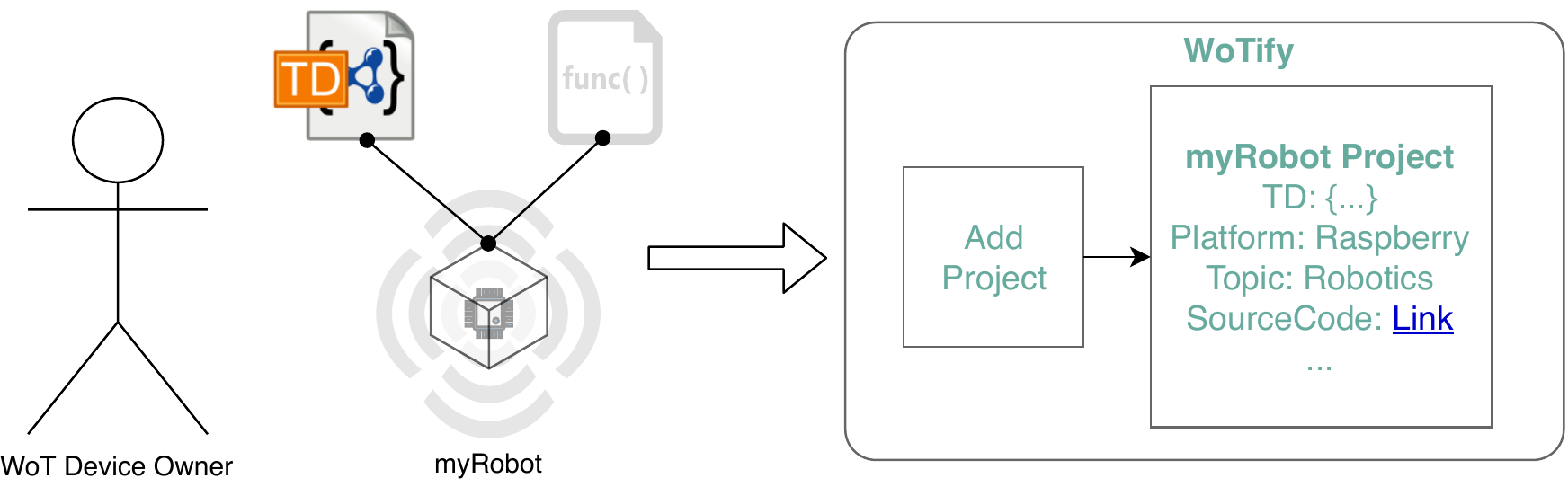}
  \centering
  \caption{How to share the implementation of a W3C WoT device on WoTify}
  \label{fig:contributeFlow}
 \end{figure}

\textbf{Contributing to WoTify library:} If one has a W3C WoT compatible device and its source code to share with the community, it can be done through WoTify. 
As shown in Figure~\ref{fig:contributeFlow}, a W3C WoT device owner:

\begin{enumerate}

  \item goes to WoTify homepage to add a new project
  \item fills in the details of the project, such as a name, a TD template, platform such Raspberry Pi, Arduino, etc., topics such as sensor, lighting, robotics and source code repository link if the device can run 3\textsuperscript{rd} party software.
  \item the project goes online, available for the community.

\end{enumerate}

In the WoT community, there are already contributors with W3C WoT compatible devices as seen in the Plugfests at Face-to-Face meetings of the working group.
These devices are built with the knowledge and expertise gained from the standardization process and would be considered as contributions to the community in general.
With the WoTify becoming online, these could be some of the first contributions that further demonstrate the use case of WoTify.

\subsection{WoTify Command Line Interface} \label{subsec:cli}

Since the beginning of 2010s, package managers and command line interfaces (CLI) for programming languages have become the defacto standard for installing different libraries.
For example, we see \emph{npm} for JavaScript, \emph{pip} for Python, \emph{Maven} for Java and even more when we consider other programming languages.
Each of these tools can browse repositories, each containing more than 100.000 projects.
Similarly, Linux distributions support package managers to install packages from different programming languages.
In all the cases, these package managers and command line interfaces, together with the vast array of different packages and projects hosted in a repository become a very strong selling point, or even sometimes referred as the \emph{killer app}\cite{perlkillerapp}.


With WoTify, we have also considered having such a CLI to enable easy installation of WoTify projects.
This CLI would enable to run a command like \texttt{wotify install myRobot-flask} which would install the project called \verb|myRobot-flask|, found at WoTify platform, on a Raspberry Pi computer.
There are two main differences compared to existing package managers and CLIs when installing a project on a device with WoTify:
\begin{enumerate}
  \item The programming language can be different for each project. This results in the need to have a wrapper for the language of the project.
  \item The project might need to be installed on an external device. This results in requiring a wrapper for copying/flashing the source code to the target device. (Cross Compiling)
\end{enumerate}

\begin{lstlisting}[caption  = {package.json file used by \emph{npm} to install a Python package}, label=list:wot-package, style=json, aboveskip=-5mm]
{
  "name": "wot-mearmpi",
  "version": "1.0.0",
  "description": "W3C WoT interface for the MeArm Pi Robotic Arm",
  "scripts": {
    "install": "pip install -r requirements.txt"
  }
}
\end{lstlisting} 


Given that most of the languages used by the current WoT community already have CLIs, package managers and repositories, we aim to leverage them and only create a wrapper around them. 
However, all the different languages require a prerequisite set of tools for the CLIs to work.
For example, to install a package with \emph{npm}, the \emph{npm} software has to be installed, along with node.js development environment.
This means that a command like \texttt{wotify install myRobot-flask} should check and install build environment, which depends on the underlying operating system.

For WoTify CLI, we are being inspired by the way \emph{npm} handles different configurations. 
This way, WoTify would be similar to a tool that a significant number of the current WoT implementations are built with.
As seen in line 6 of Listing~\ref{list:wot-package}, \emph{npm} is able to rely on a set of predefined terms that can run any command. 
In this case, even though we are using \emph{npm}, which is for the JavaScript language, we are installing a project in the Python language by overriding the \texttt{install} command on line 6.
This is not the recommended use of \emph{npm} but we want to use a similar approach in WoTify.

\section{WoTify Implementation}\label{sec:wotify}

\begin{figure}[t]
  \frame{\includegraphics[width=0.45\textwidth]{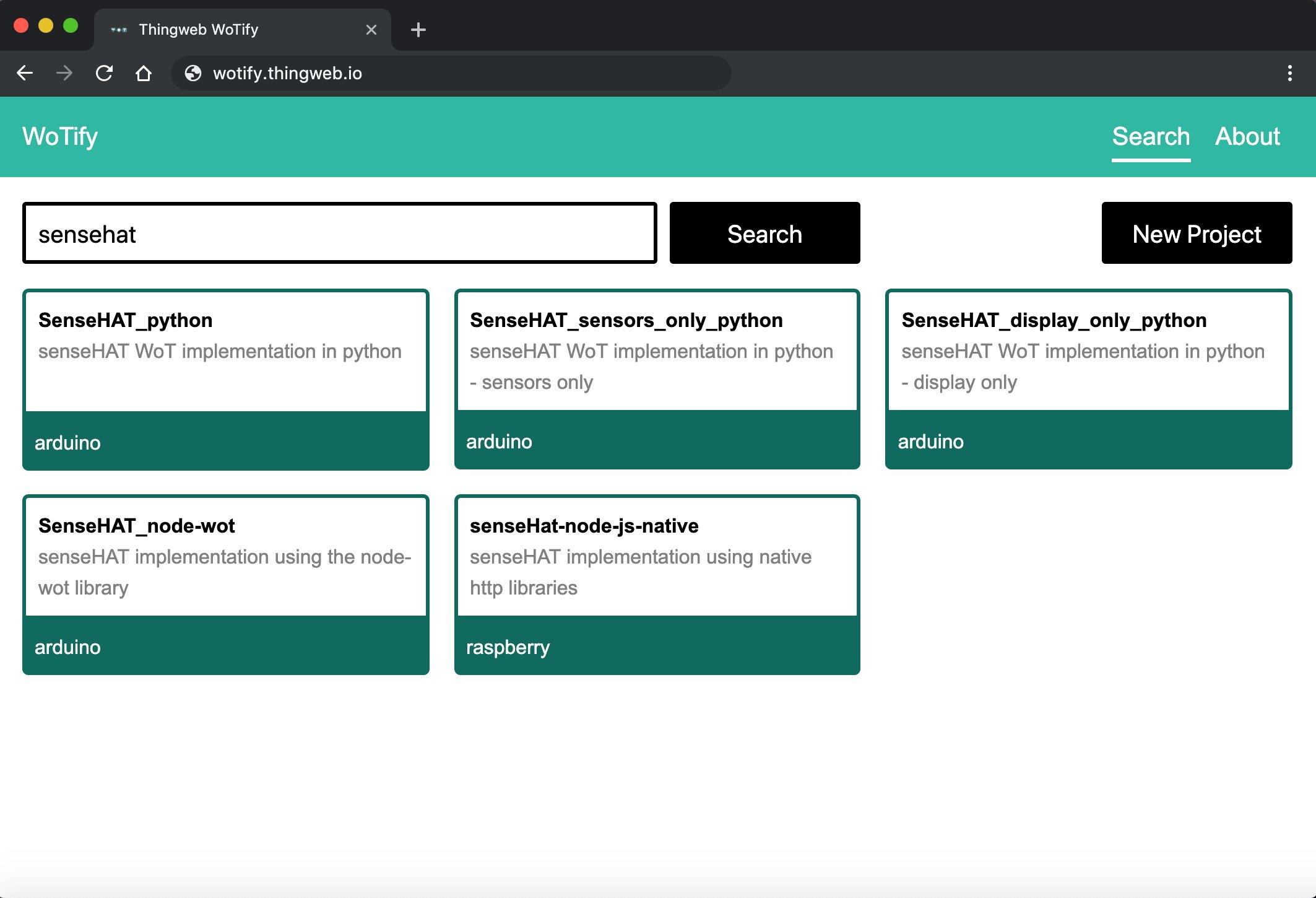}}
 \centering
   \caption{WoTify main page showing search results for Sense HAT related projects}
   \label{fig:wotify_search}
 \end{figure}

 \begin{figure}[t]
  \frame{\includegraphics[width=0.45\textwidth]{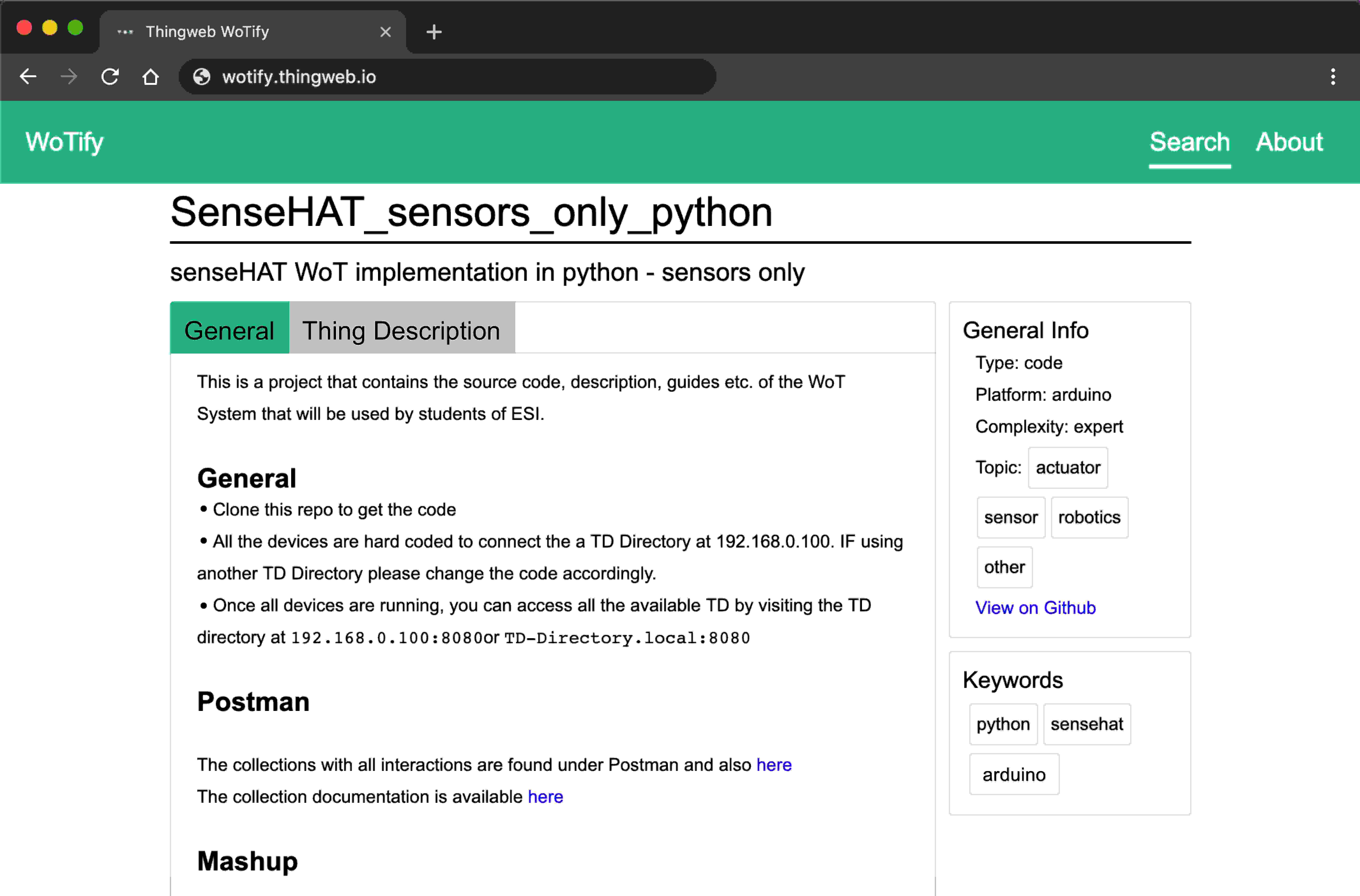}}
 \centering
   \caption{A project page on WoTify showing general information on how to use the project.}
   \label{fig:wotify_general}

 \end{figure}

 \begin{figure}[h]
  \frame{\includegraphics[width=0.45\textwidth]{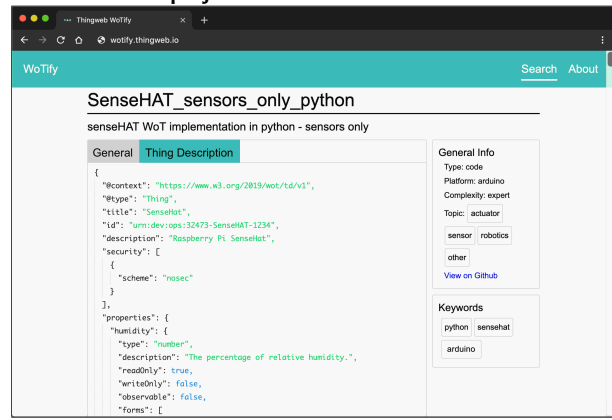}}
 \centering
  \caption{Project page on WoTify showing a Thing Description Template. This Thing Description can be downloaded in case of already existing non-W3C WoT devices and to enable interoperability with other W3C WoT devices.}
  \label{fig:wotify_td}
  \end{figure}


The concept that we have introduced in Section~\ref{sec:concept} is implemented as a Web application that will be available on the Eclipse Thingweb Homepage\footnote{\url{http://www.thingweb.io/}}.
It consists of a front-end Web user interface, a back end and a database, as illustrated in Figure~\ref{fig:wot-concept}. 
The WoTify front-end homepage allows querying and displaying available WoT projects, and allows users to contribute their own projects. 
The back end is connected to the front end and database and it handles queries, validates, retrieves and stores projects and organizes user authentication. 
The database stores all user projects with their attached data.

\subsection{Searching for a WoT Project}
As the main purposes of WoTify are the discovery of WoT projects and the contribution of these, the landing page offers searching existing projects and adding new projects. 
Users can search for a specific term, an implementation platform (e.g. Raspberry Pi), a topic or custom keywords. 
The results of an entered query will then be shown below the search bar of the homepage as shown in Figure~\ref{fig:wotify_search}. 
To better distinguish between projects that are TD templates and actual WoT implementations, the search results will be displayed in a different colors. 
When a queried project is selected by the user via clicking on it, a new page comprised of the project's details will be loaded as in Figure~\ref{fig:wotify_general}. 
Besides the standard information like name, short description and implementation type, the associated keywords, topics, level of complexity and link to Github repository can be also examined. 
However, the focus of the project page lies on the \emph{General} and \emph{Thing Description} area at the center:
\begin{itemize}
  \item The \emph{General} tab will display the content of the associated repository's main markdown file that is usually referred to as the \emph{Readme} document. 
  Contributors are strongly encouraged to describe the implementation in detail in this file and provide installation instructions and requirements. 
  If no such file is available, the project's description is viewed instead. 
  \item The \emph{Thing Description} tab displays the project's TD file which will be viewed in JSON format as shown in Figure~\ref{fig:wotify_td}.
  If the projects is meant to WoTify an already existing IoT device, this TD will be used as the template.
\end{itemize}

\begin{figure}[t]
  \frame{\includegraphics[width=0.45\textwidth]{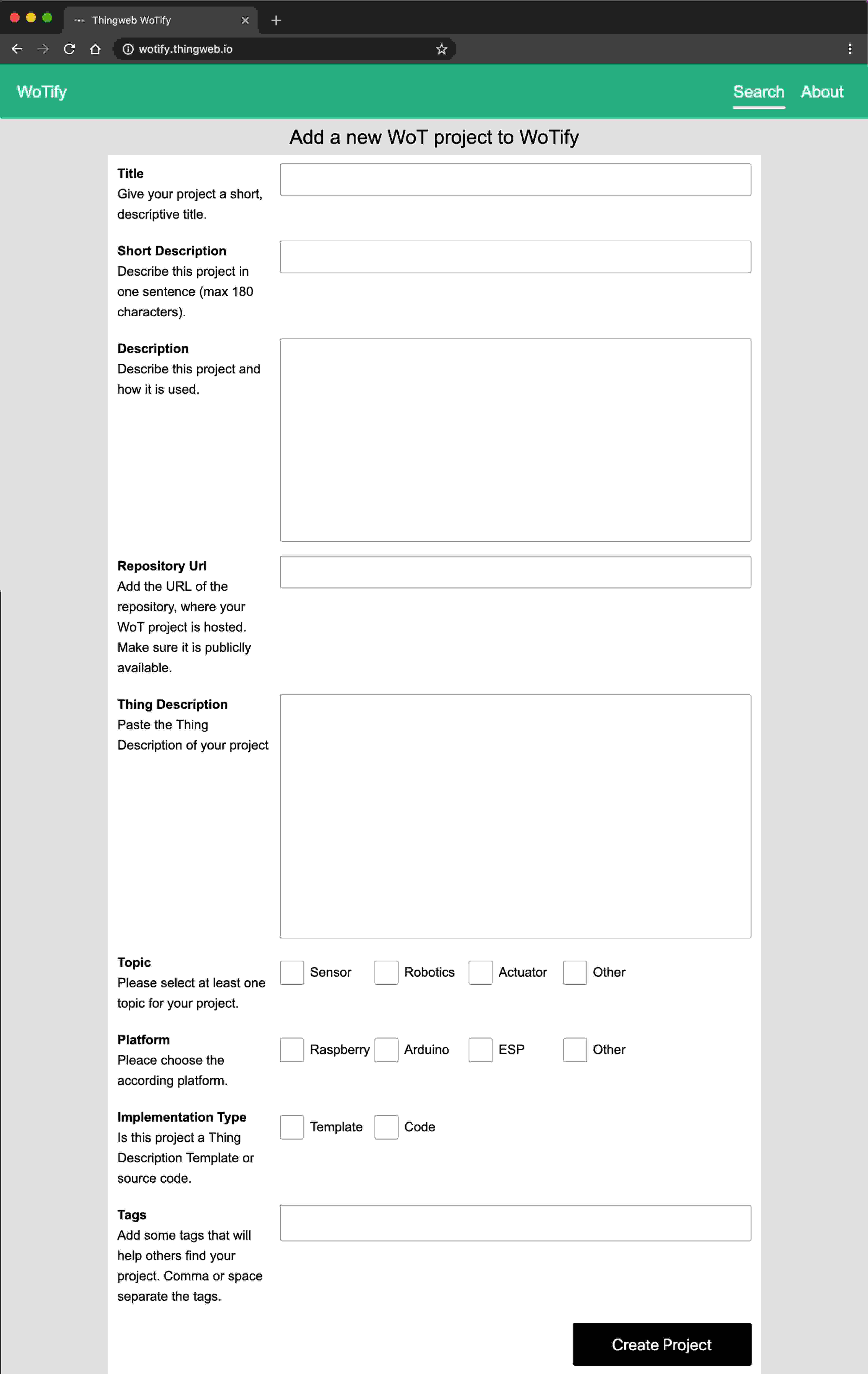}}
 \centering
   \caption{WoTify \emph{Add Project} page showing what information one has to provide to share their project on WoTify}
   \label{fig:wotify_add_project}
 \end{figure}

\subsection{Contributing to WoTify with a new Project}
If a user wants to contribute to the WoTify platform, the \emph{Add Project} button on the main page needs to be clicked. 
A form, with all the required inputs needed for a WoTify implementation will be displayed and needs to be filled out. 
This form can be seen in Figure~\ref{fig:wotify_add_project}.
To fill out the Thing Description input, the user can either paste the raw content or upload a JSON file from the local machine. 
When all required fields are correctly filled and validated by the back end, a new project will be added to the system's database and can then be retrieved by other users.

The form we use to ask for the input relies on the WoTify implementation schema that can be seen in Listing~\ref{list:wot-schema}. 

\begin{lstlisting}[caption  = {The JSON Schema WoTify uses to describe and validate the project information}, label=list:wot-schema, style=json]
  {
   title: "WoTify JSON Schema for checking implementation inputs",
   type: "object","required: [...],
   properties: {
    name: {type: "string",minLength: 5},
    shortDescription: {
     type: "string",minLength: 5,maxLength: 180},
    longDescription: {
     type: "string",minLength: 5,maxLength: 500},
    github: {type: "string",format: "uri"},
    readme: {type: "string",format: "uri"},
    implementationType: {enum:["template", "code"]},
    topic": {
     type: "array",additionalItems: false,
     minItems: 1,uniqueItems: true,
     items: {
      enum: ["sensor", "actuator", "robotics", "lighting","other"]
     }
    },
    platform:{enum:["raspberry","arduino","ESP","other"]},
    tags: {
     type:"array",additionalItems:false,minItems:1,
     uniqueItems: true,items: {type: "string"}
    },
    complexity:{enum:["simple","medium","expert"]},
    version: {
     type: "object",required: ["instance"],
     properties:{instance:{type:"string"}}
    },
    td:{type:object,properties:{...},required:[...]}
   }    
  }
  \end{lstlisting} 

\textbf{Used Technologies:} The back end is based on Node.js, the front end is implemented using the Vue.Js frontend framework. 
MongoDB was chosen as database, as it is based on the JSON format and therefore aligns with JSON based WoT technologies. 
These implementation choices allow our system to be highly extensible and scalable. 

\section{Related Work}\label{sec:sota}

In recent years, various concepts have been developed to integrate different IoT offerings (e.g. services such as properties or actions from physical Things) into existing IoT ecosystems. 
The BIG IoT (Bridging the Interoperability Gap of the Internet of Things) project~\footnote{\url{http://big-iot.eu/}} developed a marketplace on which IoT offerings can be setup, discovered and used  through simple integration into new IoT services or applications\cite{Broring2017}.  
As a technical concept, the BIG IoT ecosystem is based on a so-called Offering Description (OD), which can be seen as a similar approach to the W3C Thing Description. 
WoTify can be also seen as a marketplace that offers (existing) WoT projects for the community. 
However, WoTify is independent of a specific IoT ecosystem such as BIG IoT and uses the upcoming standardized W3C WoT building blocks to make services or devices WoT enabled.

We also see Thingweb Directory\footnote{\url{https://github.com/thingweb/thingweb-directory}} as an important work towards hosting WoT related projects. 
Thingweb Directory is designed to host TDs and offer semantic search to find TDs of (once) running devices.
It is similar to WoTify, as WoTify can also host TDs, but the main difference is that WoTify is not limited to TDs and its aim is not to find TDs of other devices but to turn any device into a WoT device.

\section{Outlook}\label{sec:outlook}

The task we have overtaken with WoTify is still in its infancy regarding the features it can offer based on the concept introduced in Section~\ref{sec:concept}.
We are planning to continue adding new features and shape it according to the feedback from the community.
We have already seen the following features as the most needed in the short run:
\begin{itemize}
  \item \textbf{WoTify CLI:} The WoTify CLI has been introduced in Section~\ref{subsec:cli}, but this part is not ready yet. 
  Thus, in the current state of WoTify, developers need to install the WoTify projects according to the General Information page of the project.
  \item \textbf{TD Template Editor:} In the current state of WoTify, TD templates have to be edited manually after downloading them.
  This can be an error-prone process with errors resulting in a cumbersome debugging process since one would need to read the API description of the given device. 
  An embedded TD editor that forces the developer to input only the relevant and correct information is necessary in the development of WoTify.
  \item \textbf{Rating System:} It is very important to stress the fact that WoTify projects that are downloaded will run on actual physical devices.
  Since we cannot run every WoTify project for each intended device, the community will need to be involved in a certain feedback mechanism to ensure the quality of the projects.
  In platforms like \emph{npm}, this can be shown by weekly download counts or in Node-RED Library, this is done by users giving stars as rating.
  This feature would also help to show the support of the community, encouraging the adoption of W3C WoT.
\end{itemize}



Additionally, the work of the W3C WoT Testing Task Force can be used in the future to certify the correctness of the projects, strengthening the trust on projects.
We can foresee tools like WoT Test Bench\footnote{\url{https://github.com/tum-ei-esi/testbench}} integrated into WoTify CLI to run tests before publishing the project. 
This verification can be further enhanced by the contributors when they provide the recently introduced path descriptions\cite{Korkan2018}.
The paths would also provide the developers a more constrained but safer way of using WoT devices.

\section{Conclusion}

In this paper, we have presented a new platform for W3C Web of Things projects that is already online for the WoT community.
This platform is called WoTify and it is able to host WoT projects that can be used to WoTify existing devices, including brownfield IoT devices.
Additionally, we have proposed a new type of command line interface for managing WoTify projects that is independent of programming languages.
We believe that thanks to WoTify, the W3C WoT can be adopted by a wide array of developers, allowing them to quickly deploy the Web layer on various Internet connected devices and turn them into W3C WoT compatible devices.

\vfill

\vspace*{-0.3em}

\renewcommand{\baselinestretch}{1.15}
\bibliographystyle{ACM-Reference-Format}
\bibliography{references}


\end{document}